\documentclass{an}
\usepackage{times,graphicx,anlongtable,ngerman}
\begin{document}
\Pagespan{0}{}
\Yearpublication{Year of publication}
\Yearsubmission{Year of submission}
\Month{Month of publication}  
\Volume{Volume}
\Issue{Issue}
\DOI{DOI}
\title{New brown dwarf candidates in the Pleiades
 \thanks{Based on observations obtained with telescopes of the University Observatory
Jena, which is operated by the Astrophysical Institute of the Friedrich-
Schiller-University.}}
\author{
T. Eisenbeiss\inst{1} \thanks{E-mail: eisen@astro.uni-jena.de}
	\and M. Moualla\inst{1}
	\and M. Mugrauer\inst{1}
	\and T.~O.~B. Schmidt\inst{1}
	\and St. Raetz\inst{1}
	\and R. Neuh\"auser\inst{1}
	\and Ch. Ginski\inst{1}
	\and M. M. Hohle\inst{1,2}
	\and A. Koeltzsch\inst{1}
	\and C. Marka\inst{1}
	\and W. Rammo\inst{1}
	\and A. Reithe\inst{1}
	\and T. Roell\inst{1}
	\and M. Va{\v n}ko\inst{1}
	}
\institute{
	Astrophysikalisches Institut und Universit\"ats-Sternwarte Jena, Schillerg\"asschen 2-3, 07745 Jena, Germany \and
	Max Planck Institut f\"ur extraterrestrische Physik Garching, Giessenbachstrasse, 85738 Garching
	}
\received{R-date}
\accepted{A-date}
\publonline{P-date}
\abstract{We have performed deep, wide-field imaging on a $\sim0.4\,$deg$^2$ field in the Pleiades (Melotte 22). The selected field was not yet target of a deep search for low mass stars and brown dwarfs. Our limiting magnitudes  are $R\sim22$\,mag and $I\sim20$\,mag, sufficient to detect brown dwarf candidates down to $40\,M_{J}$. We found 197 objects, whose location in the ($I$,$R-I$) color magnitude diagram is consistent with the age and the distance of the Pleiades. Using CTK $R$ and $I$ as well as JHK photometry from our data and the 2MASS survey we were able to identify 7 new brown dwarf candidates. We present our data reduction technique, which enables us to resample, calibrate, and co-add many images by just two steps. We estimate the interstellar extinction and the spectral type from our optical and the NIR data using a 2-dimensional $\chi^2$ fitting.}
\maketitle

\section{Introduction}
Imaging surveys for brown dwarfs (BDs) have often targeted young open clusters or star-forming regions because their ages and distances are known and because BDs are brighter and warmer when they are young (e.g. Bouvier et al., 1998; Moraux et al., 2003). The first BD in the Pleiades, PPl\,15 was detected by Basri, Marcy, \& Graham (1996) applying the lithium test method. Soon afterward the BDs Teide\,1 and Calar\,3 (Rebolo et al. 1996) were detected. According to the number of published papers the Pleiades cluster is the preferred ground for hunting brown dwarfs due to its nearly ideal properties. Those properties are: (a) its age of about 119\,Myr (Mart\'in \& Dahm, 2001), where objects at the hydrogen burning mass limit (HBML) are still relatively warm; (b) its distance of about 135.5\,pc (An et al., 2007) which is near enough that the apparent magnitude of HBML objects is bright enough to be observed even with relatively small telescopes; (c) its richness, with about 2100 known stars and brown dwarfs (Mermilliod, 1998) and (d) its relatively high galactic latitude, which minimizes the number of background objects.

We selected a field of the Pleiades and did deep imaging in two bands to detect faint, red objects which could be brown dwarfs or low mass stars. We describe our observations in section \ref{obs}. Data reduction is described in section \ref{data}, photometric analysis is given in section \ref{photo}. The $\chi^2$ fitting algorithm of the interstellar extinction and resulting candidates are presented in section \ref{res} and discussed in section \ref{concl}.

\section{Observations}\label{obs}
\begin{figure*}
\resizebox{\hsize}{!}{\includegraphics{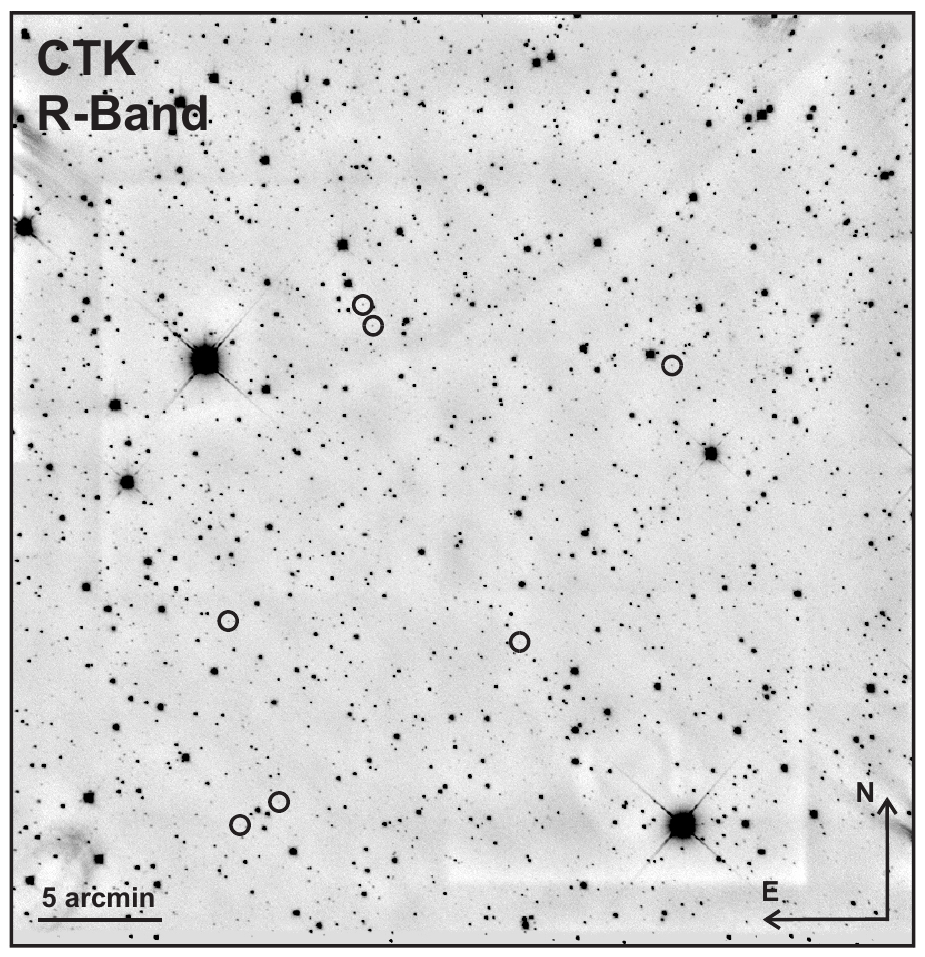}}
\caption{Our observed field. The summed up $R$ Filter image with a total exposure time of 36\,ks is shown. The center of the field is $\alpha=3^h42^m20.6^s$, $\delta=+25^{\circ}36'54''$. North is up east is left. The brown dwarf candidates we have detected are marked by circles, see also Fig. \ref{cut3} to Fig. \ref{cut8}. The illumination effects at the edges of the image are reflections of bright stars. The total covered field in this image is about $36.5 \times 37.7$ arcmin$^2$. The two bright and saturated stars are the F5 star HD\,23075 (NE) and the K2 star HD\,22915.}
\label{Rim}
\end{figure*}

\begin{figure*}
\resizebox{\hsize}{!}{\includegraphics{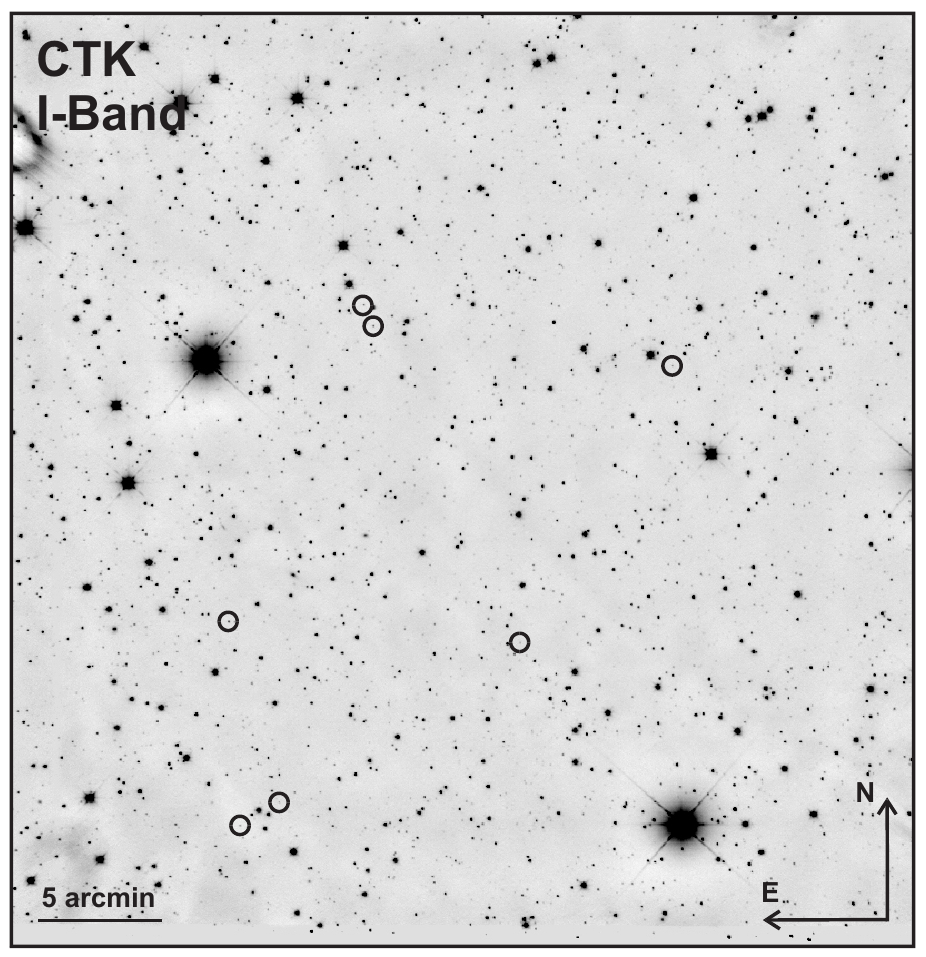}}
\caption{The summed up $I$ Filter image with a total exposure time of 47\,ks is shown. The center of the field is $\alpha=3^h42^m20.6^s$, $\delta=+25^{\circ}36'54''$. North is up east is left. The brown dwarf candidates we have detected are marked by circles. For more details see Fig. \ref{Rim} and Fig. \ref{cut3} to Fig. \ref{cut8}.}
\label{Iim}
\end{figure*}

Starting in March 2007 we have observed a field at the edge of the Pleiades cluster, which was not a target of extended surveys before (see Tbl. 1 in Schwartz, \& Becklin 2005). The field is located at $\alpha=3^h42^m20.6^s$, $\delta=+25^{\circ}36'54''$. For these observations we have used the Cas\-se\-grain-Te\-le\-skop-Ka\-me\-ra (CTK) 
{which is installed at the 250\,mm Cas\-se\-grain auxiliary telescope of the 900\,mm reflector} 
which is operated at University Observatory Jena.
The instrumental details are described in Mugrauer (2009). With its $1024\times 1024$ Pixel CCD detector  and a resolution of about 2.2\,arcsec/pixel the CTK covers a field of view (FoV) of $\sim 38' \times 38'$. In total we have collected 599 images in $R$ and 781 images in $I$ which yields to a total integration time  36\,ks and 47\,ks respectively (Fig. \ref{Rim} and Fig. \ref{Iim}), see Tbl. \ref{obslog}. 
\begin{table}
\caption{Observations log}
\label{obslog}
\begin{tabular}{lccc}
\hline
\hline
Date & Filter & tot. Exposure \\
     & Bessel& [$s$]&\\
\hline
March 11th, 2007 & $R$ & 5400 \\
March 12th, 2007 & $R$ & 5280 \\
March 13th, 2007 & $R$ & 5520 \\
March 14th, 2007 & $R$ & 4380 \\
March 15th, 2007 & $R$ & 4380 \\
Jan  8th, 2008 & $R$ & 3540 \\
Jan  9th, 2008 & $R$ & 1140 \\
Jan 13th, 2008 & $R$& 2340 \\
March 23rd, 2008 & $R$ & 1800 \\
March 29th, 2008 & $R$ & 1620 \\
\hline
Sep 26th, 2008 & $I$ & 6480 \\
Sep 27th, 2008 & $I$ &  720 \\
Sep 28th, 2008 & $I$ & 9120 \\
Sep 29th, 2008 & $I$ & 6960 \\
Oct  7th, 2008 & $I$ & 3540 \\
Oct 12th, 2008 & $I$ & 3060 \\
Oct 23rd, 2008 & $I$ &11520 \\
Oct 24th, 2008 & $I$ & 1320 \\
Oct 26th, 2008 & $I$ & 3660 \\
\hline
\end{tabular}
\end{table}
The majority of the $R$ band images was taken at high airmasses, so the photometric quality of these data is lower than that of the $I$ band data. Nevertheless, the quantum efficiency of the detector is better in $R$ than in $I$. {Considering this, the different observing conditions in both observation campaigns ($R$ and $I$), and regarding our special interest in red objects the total integration time gives a comparable sensitivity in the co-added images.}

\section{Data reduction}\label{data}
The individual integration time per image was 60\,s in $R$ and $I$, because we want to study both, faint objects as well as their photometric variability. The latter is still under investigation and will be presented later elsewhere (Moualla et al. in prep).

All images were flat fielded and dark subtracted using the ESO software \textsl{ECLIPSE} (Devillard 2001). A bad pixel mask calculated from the sky flat as well as a 3$\sigma$ filtering algorithm were used to clean the images from bad or hot pixels.  In order to co-add all images in just a few steps we used the \textsl{TERAPIX}\footnote{Traitement elmentaire, Reduction et Analyse des PIXels de megacam} (Bertin et al. 2002) software \textsl{SCAMP}\footnote{Software for Calibrating AstroMetry and Photometry} (Bertin 2006).  The calculations of \textsl{SCAMP} are based on catalogs produced by \textsl{Source Extractor} (\textsl{SE}, Bertin \& Arnout 1996). {The \textsl{SE} uses thresholding and deblending\footnote{For a detailed description of the source detection algorithm used by SE see Bertin \& Arnout (1996) and references therein.} for source detection, which is in fact more convenient for galaxies than for stars. However the limiting factor for the astrometric accuracy is undersampling, given the CTK pixel scale.} Using such object catalogs for each image and the 2MASS catalog (Cutrie et al. 2003) obtained from {ViZier} (Ochsenbein et al. 2000) as astrometric reference \textsl{SCAMP} computes astrometric solutions using a $\chi^2$ algorithm and applies a polynomial based field distortion correction, involving distortion coefficients up to an order of 10.

\begin{figure*}
\resizebox{1.0\hsize}{!}{\includegraphics[angle=270]{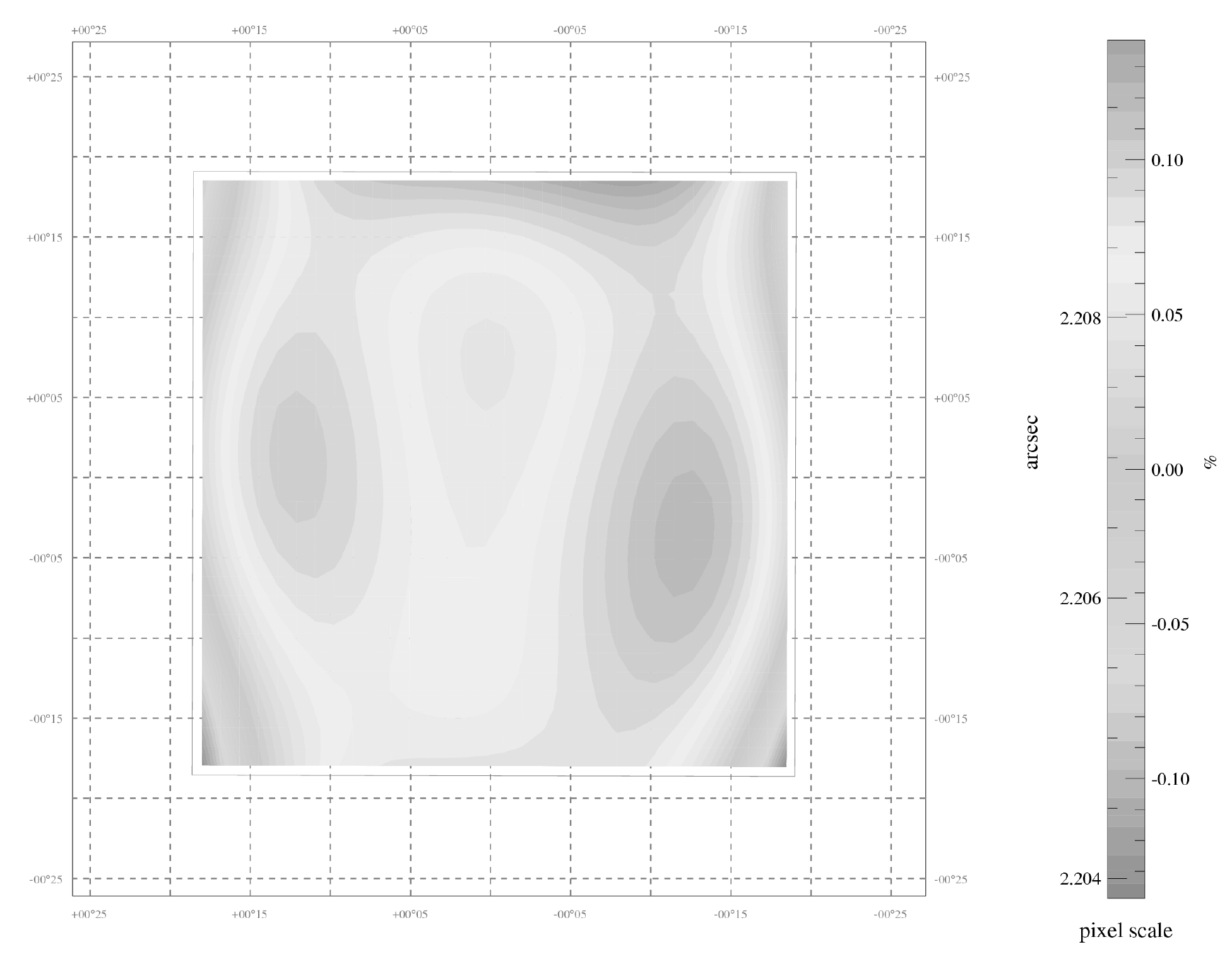}}
\caption{Field distortion map of the CTK detector in October 2008. This is originally a colored image. 
Color coding goes from red (dark spots in the image and top of the scale) through yellow and green  (light spot in the middle and vertical light stripes) to blue (corners of the image and bottom of the scale.) The pixel scale variations are given at the color bar, left in arcsec/pixel, right in \%. The difference between the center and the edge is about 0.2\%, hence, less than 5 mas.}
\label{distort}
\end{figure*}

We calculated the astrometric solution for each image and therefore the field distortion pattern of the CTK with a polynomial of the fifth order, shown in Fig. \ref{distort}. Given those distortion corrections we co-added all images of each filter using the software SWarp (Bertin et al. 2002). Subtracting all objects, detected by the SE SWarp applies a mesh based background algorithm to estimate the background. Since we got small scale illumination effects in the image, produced by bright nearby stars outside the field we used a mesh size of just 32 pixel to overcome these effects. Nevertheless the region around saturated stars is excluded from analysis anyway, since the background flux is overestimated around these regions, see Fig. \ref{Rim} and Fig. \ref{Iim}.

\section{Photometry}\label{photo}
We tried different photometry approaches. We achieved best results, hence reasonable errors and an accurate detection of faint objects, by using the source extractor for the source detection and ESO \textsl{MIDAS}s aperture photometry program (see e.g. Banse et al. 1987) for obtaining the magnitudes. The SE uses thresholding and deblending to distinguish between objects and background and to disentangle close objects whose illumination patterns are mixed. In that way also faintest objects are detectable. 

We determined instrumental magnitudes
for all sources in the co-added image
found by SE using aperture photometry
with the \textsl{MIDAS} command magnitude/circle;
the aperture radius used in our image 
for the source photons is 4.5 pixels,
the background was measured in a ring
around the source with radii
between 15 and 22 pixels.
\begin{table*}
\caption{Known Pleiades in the investigated field with known $R$ and $I$ band photometry and with membership probability larger than 0.8 (Deacon \& Hamply 2004a). (There is also a known Pleiades brown dwarf at $\alpha=03:41:41$, $\delta=+25:54:23$ which happens to lie a few arcseconds outside our field.)}
\label{cal}
\begin{center}
\begin{tabular}{l c@{\ }c@{\ }c c@{\ }c@{\ }c c c c}
\hline\hline
Id & \multicolumn{3}{c}{$\alpha$ (J2000)} & \multicolumn{3}{c}{$\delta$ (J2000)} & Rmag &Imag& prop.\\
\hline
Cl* Melotte 22 HHJ 270 &  3&42&8.270 & 25&37&0.06 & 16.831 & 14.985 & 0.875 \\
Cl* Melotte 22 SRS 82788 &  3&42&4.850 & 25&39&48.01 & 14.716 & 13.326 & 0.895 \\
Cl* Melotte 22 SRS 79717 &  3&42&43.800 & 25&32&6.09 & 15.847 & 14.479 & 0.905 \\
Cl* Melotte 22 HCG 55 &  3&41&54.200 & 25&43&47.08 & 15.893 & 14.444 & 0.923 \\
Cl* Melotte 22 HCG 96 &  3&43&7.560 & 25&34&29.07 & 16.763 & 14.908 & 0.955 \\
Cl* Melotte 22 MSH 9 &  3&42&3.410 & 25&22&39.08 & 17.235 & 15.442 & 0.890 \\
Cl* Melotte 22 MBSC 19 &  3&43&34.130 & 25&35&26.06 & 16.072 & 14.454 & 0.957 \\
Cl* Melotte 22 SK 733 &  3&41&10.260 & 25&45&56.07 & 16.503 & 14.790 & 0.942 \\
Cl* Melotte 22 MBSC 22 &  3&43&36.660 & 25&47&1.03 & 16.172 & 14.441 & 0.872 \\
Cl* Melotte 22 SRS 76506 &  3&43&35.200 & 25&24&31.06 & 15.998 & 14.426 & 0.947 \\
Cl* Melotte 22 SRS 76505  &  3&43&37.310 & 25&24&32.08 & 15.396 & 14.035 & 0.763 \\
\hline
\end{tabular}
\end{center}
\end{table*}

As reference for photometric calibration we have used the catalog of Deacon \& Hamply (2004a) published in Deacon \& Hamply (2004b). They measured $RI$ photometry and proper motion of stars in this region of the Pleiades cluster from Schmidt plates and calculated membership probabilities. In total 15 stars of our field where available. These stars are partly known as flare stars and the accuracy of the Schmidt plates photometry decreases for bright objects. So we excluded 4 of the 15 stars and used the remaining 11 (given in Tbl. \ref{cal}) to calibrate our images. {The zero point was defined as difference between our instrumental magnitudes and that reported by Deacon \& Hamply (2004a). Taking the mean of these differences of all 11 stars} we get zero points of
$C_R=(-19.26\pm 0.26)\,$mag and
$C_I=(-18.52\pm 0.17)$\,mag for the $I$ band image respectively. {We give the 1-$\sigma$ calibration error ($\sigma_{cal,R}=0.26$, $\sigma_{cal,I}=0.17$). The typical accuracy of our measurements is $\sigma_{mesh,R}=0.18$ and $\sigma_{mesh,I}=0.14$ .
The individual integration time was 60s per image.}
Objects with apparent magnitudes down to $\sim22\,$mag in $R$ and $\sim 20$\,mag in $I$ are  detectable in the average of all our CTK images taken in R- and I-band.

\section{Results}\label{res}
In total we detected 2401 objects in $R$ and $I$ (only objects visible in both, the $R$ and the $I$ image, are studied further). The objects are {cross-correlated with} each other using a self written procedure originally designed to calculate exact proper motions (Eisenbeiss et al. 2007). {The objects are detected at a threshold of 1.5 (relative to the sky) and with 64 deblending grades (maximum for the SE). Since there are small scale background fluctuations in the image we applied a mesh based background calculation with a mesh size of 16 pixels only.} We obtained $R$$I$ photometry for known Pleiades members from the WEBDA database (Mermilliod 1998). We used basic $R$$I$ Johnson photometry for the bright stars and a list of CCD Cousins observations of faint objects, given at WEBDA. We can now plot our objects in a $R-I$ vs. $I$ color magnitude diagram. 

\begin{figure}
\resizebox{\hsize}{!}{\includegraphics{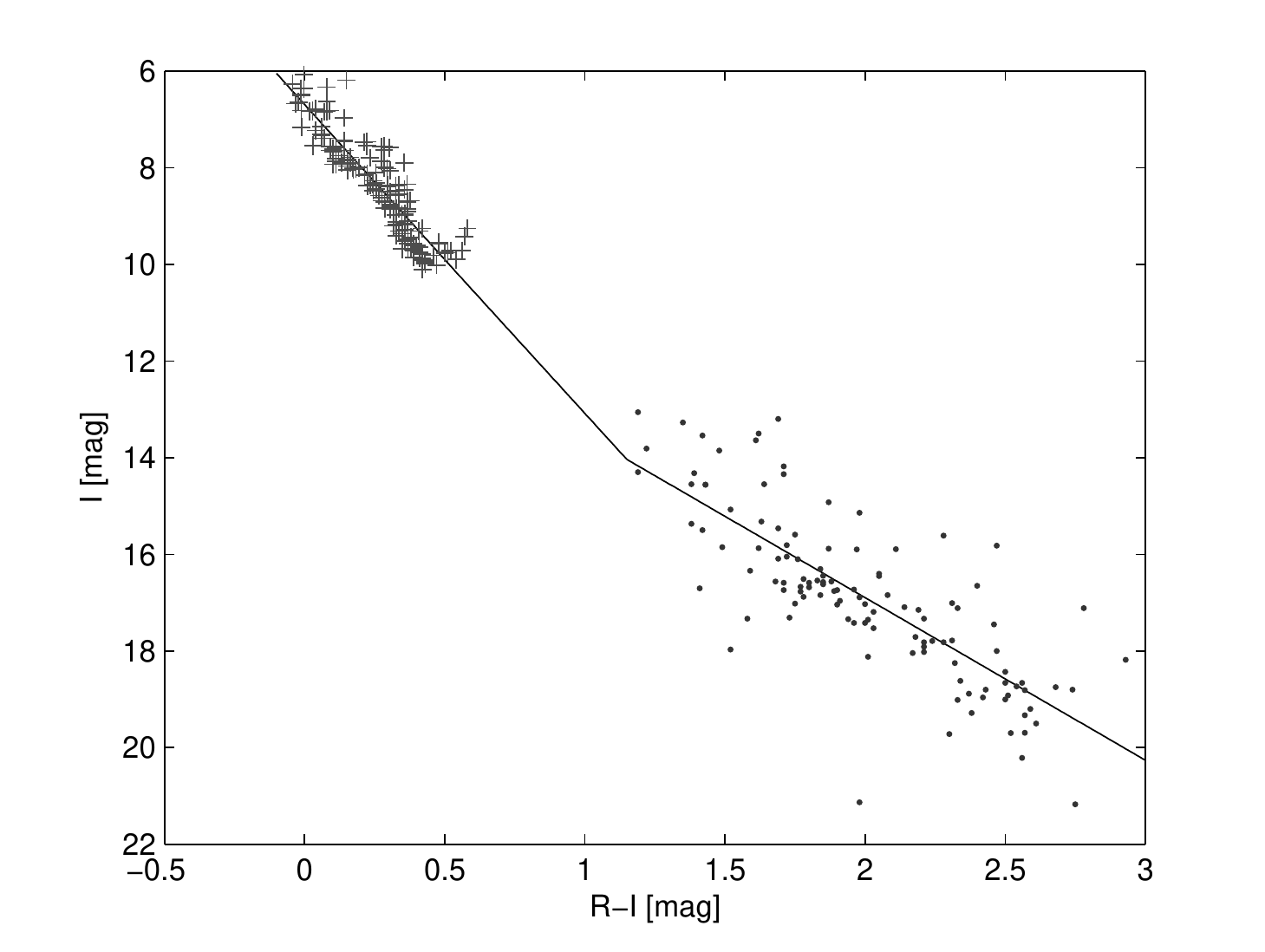}}
\caption{Color magnitude diagram $I$ vs. $R-I$. Crosses represent $RI$ Johnson photometry of known Pleiades. Dots represent $RI$ CCD Cousins photometry of known Pleiades low mass stars and brown dwarfs. The black linear slopes are the fits to the data: stellar main sequence is top left, low mass objects is bottom right.}
\label{pleical}
\end{figure}

To plot the Pleiades main sequence, we can now use data from all known Pleiades members while we were limited to known Pleiades within our FoV for the photometric calibration of the image. Both are comparable to our Bessel filter system. Based on this literature data we fitted the Pleiades by three linear slopes (Fig. \ref{pleical}). The first straight line fits the Pleiades main sequence. The third straight line fits the Pleiades brown dwarf locus and the second straight line interpolates in between these regions. {This approximation represents the Pleiades main sequence with sufficient accuracy, according to our photometric precision.} (Fig. \ref{pleical}).

\begin{figure}
\resizebox{\hsize}{!}{\includegraphics{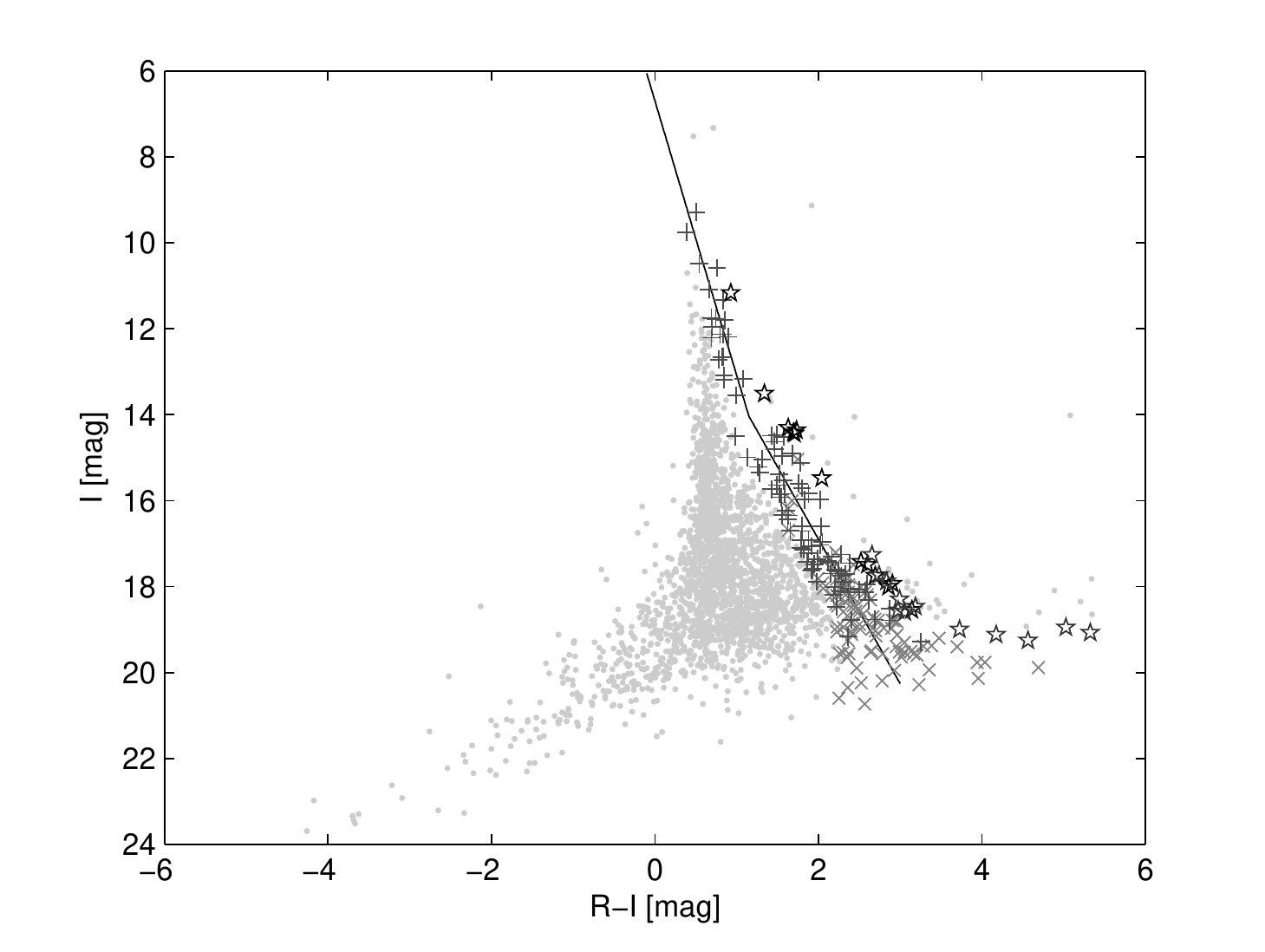}}
\caption{The light gray dots are the detected background stars in our observed field. dark gray pluses are the candidate members of the Pleiades with associated 2MASS photometry. The crosses denote the Pleiades membership candidates detected by us, but not by the Two Micron All Sky Survey. The five-pointed stars denote objects, which might be Pleiades candidates under the assumption that these sources are unresolved binaries.}
\label{pleibd}
\end{figure}

The photometry error of each source in the co-added $R$ and $I$ image is calculated through the calibration error 
$\sigma_{cal}$ 
({see section \ref{photo}}) and the accuracy of the measurement ({random error, see section \ref{photo}}) 
$\sigma_{mesh}$ as $\sigma_{phot,col}^2 = \sigma_{cal}^2 + \sigma_{mesh}^2$. 
{The photometry errors of the $R$ and $I$ measurements for each object define the error of the $R-I$ color term as $\sigma_{phot,R-I}^2 = \sigma_{phot,R}^2 + \sigma_{phot,I}^2$. 
In a $I$ vs. $R-I$ diagram the full photometry error is than defined as 
$\sigma_{phot} = \sqrt{\sigma_{phot,R-I}^2 + \sigma_{phot,I}^2}$.} 
The total offset $\Delta$ allowed for membership must not exceed the photometry error of the object plus the mean residuals $\sigma_{res}$ of the fitting \linebreak
$\Delta \leq \sqrt{\sigma_{phot}^2 + \sigma_{res}^2}$. 
{$\sigma_{res}$ is defined as the mean of the fitting residuals of all sources with respect to the approximating lines}. This is illustrated in Fig. \ref{pleibd}, {or in words: The 2D-distance of one source in the $I$ vs. $R-I$ diagram from the line fitted Pleiades main sequence must not exceed the quadratic sum of the photometry error and the fitting residuals. 

Additionally each object was checked with the method described above, but under the assumption of unresolved binarity. Such objects would be located slightly above the Pleiades main sequence, but might be additional membership candidates, since the true apparent magnitude of each single object would be higher (up to 0.75\,mag if the source was an unresolved equal mass binary). Such objects are marked with five-pointed stars in Fig. \ref{pleibd}.}

This way we detected 197 objects with $RI$ photometry consistent with the Pleiades.

\begin{figure}
\resizebox{\hsize}{!}{\includegraphics{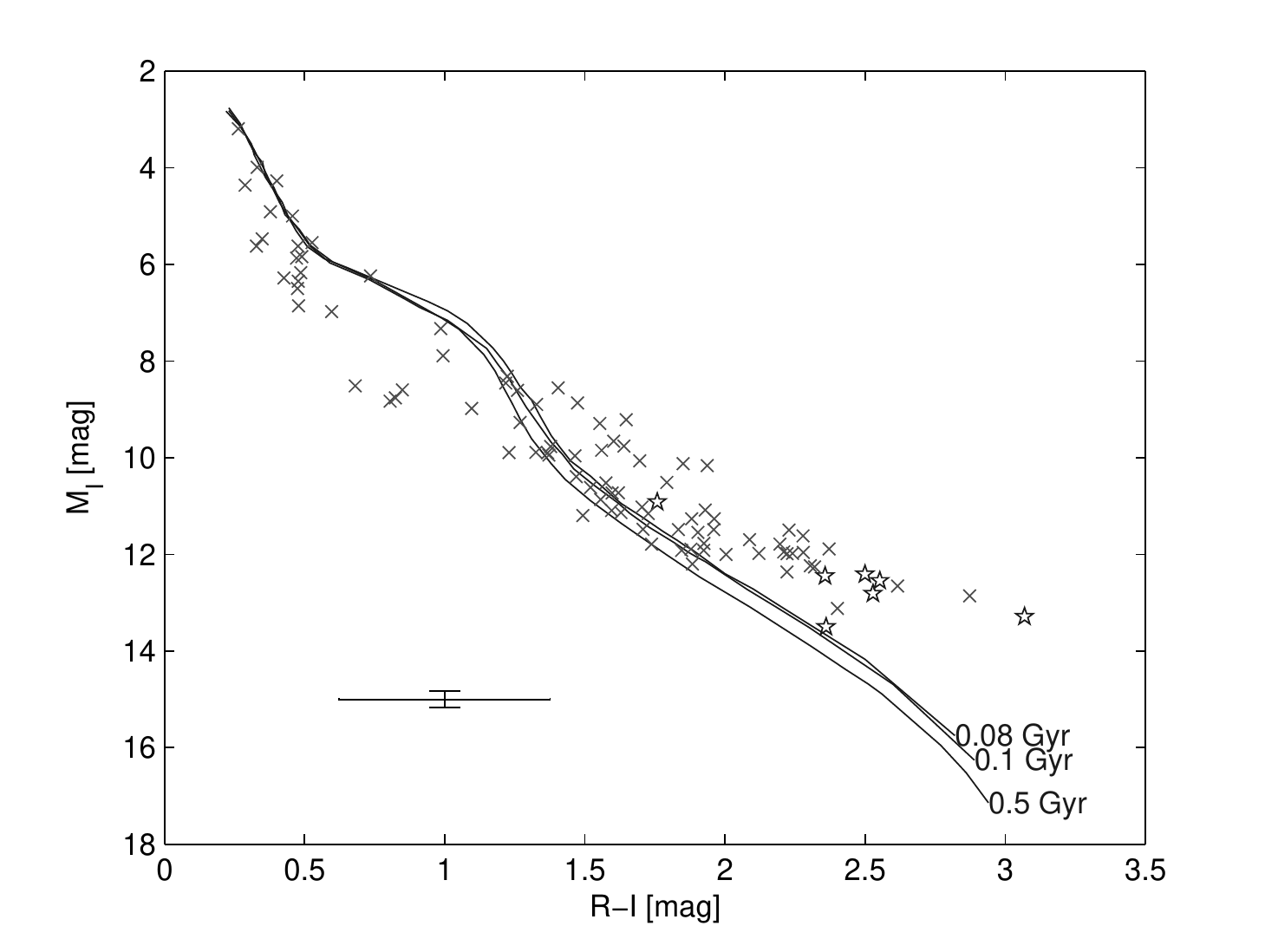}}
\caption{Comparison of the candidate members of the Pleiades (x) with three sets of theoretical isochrones. The lines are the Baraffe isochrones from 1998 (Baraffe et al. 2002) for 0.08, 0.1, and 0.5\,Gyr. A distance module of 5.66\,mag as reported by An et al. (2007) is applied. The agreement is acceptable given the uncertainties of the data and the models. The ages of the isochrones selected according to the reported age of the Pleiades of $\sim 120\,$Myr. The brown dwarf candidates found in this study are marked with five-pointed stars. For further explanations see text.}
\label{rimod}
\end{figure}

{For comparison we plotted our data with theoretical isochrones. We used the distance module given in An et al. (2007) of 5.66\,mag and plotted our Pleiades membership candidates together with one theoretical sets of isochrones Baraffe et al. (2002), see Fig. \ref{rimod}. The Baraffe models should have a good treatment of the M-dwarfs and brown dwarfs, but our objects are lying above these isochrones. Given the uncertain distance of each object, the limited photometric accuracy of our data and the theoretical difficulties in the modeling of cool dwarfs (dusty atmosphere) the agreement is acceptable.}

Then we cross checked our object list with the ''Two micron all sky survey'' catalog of point sources (2MASS, Skrutskie et al. 2006). If our sources are listed in this catalog the corresponding identifier is given in the tables. {Again we used our self written software to associate the 2MASS sources with ours. We gave an initial search radius of \linebreak 1\,arcsec $\times$ epoch difference ($\sim 10\,$yr), but the software selects always the closest source as reference if there is more than one source in the search radius. This way we avoid to miss objects, but mismatching can occur if there is no source in our images at the position of the 2MASS source. Since we are interested in the faintest 2MASS objects we did not exclude any source by having bad photometry flags.}
If the object was not found in the 2MASS catalog we named it GSHCTK (for Gro{\ss}\-schwab\-hau\-sen-Cassegrain Teleskop-Kamera) and coordinates. 

\begin{figure}
\resizebox{\hsize}{!}{\includegraphics{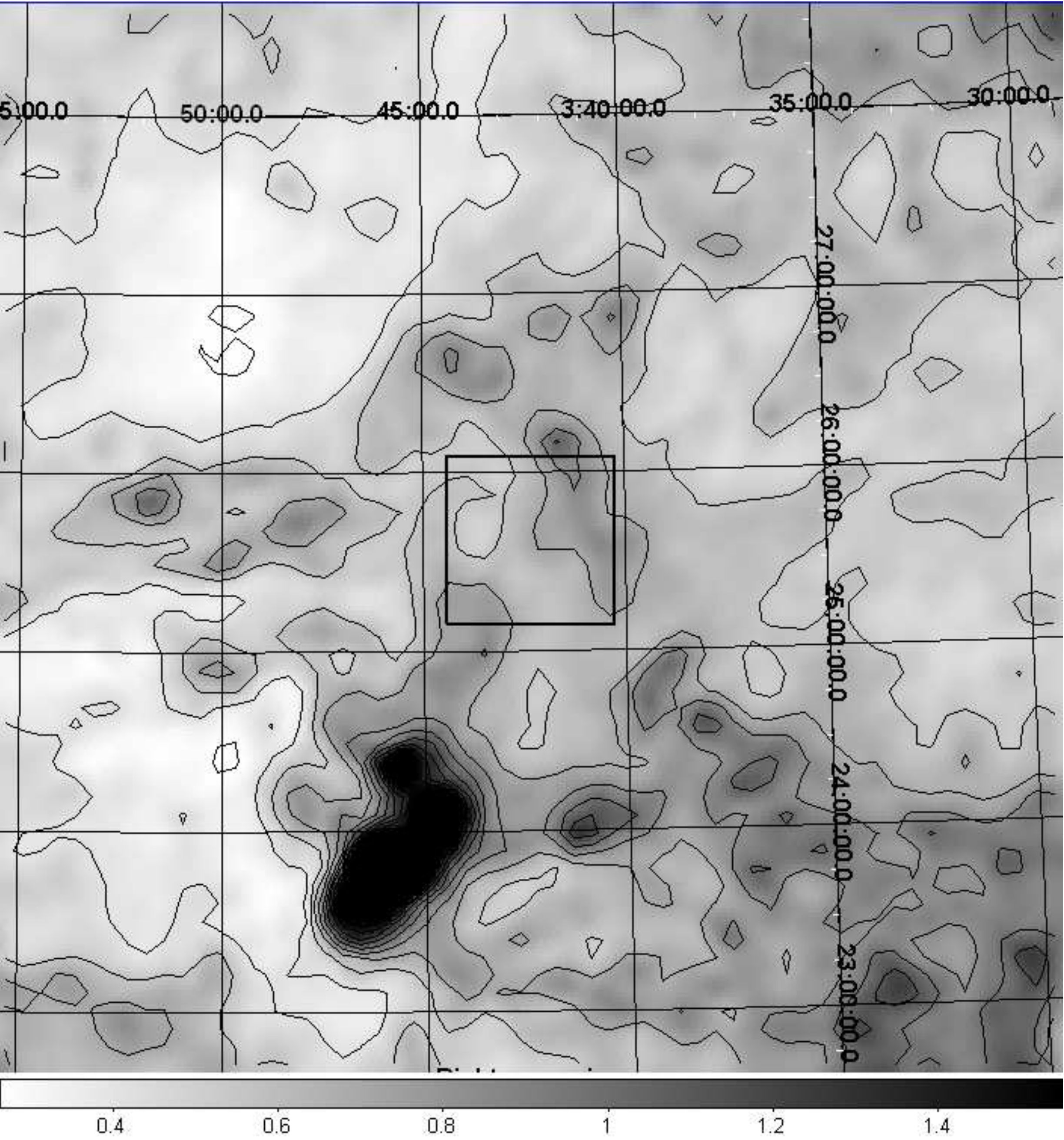}}
\caption{Extinction map of the Pleiades region, obtained from the {NAS/IPAC infrared science archive}. Our FoV is marked by a black box. The interstellar extinction in the V-Filter ($A_V$) is color coded. One can clearly see that our observations are taken in a region of low extinction.}
\label{extinct}
\end{figure}

{1244} of all sources {(Pleiades candidates and others)} are detected by the 2MASS. We estimated the spectral type\linebreak (SpT) and the interstellar extinction ($A_V$) for the sources, found in the 2MASS catalog. For these objects enough color information is available. We downloaded an interstellar extinction map from the {NAS/IPAC infrared science archive website}\footnote{http://irsa.ipac.caltech.edu/applications/DUST/} and estimated the expected extinction in the visual to be 
$A_V\approx 0.5 \ldots 1$ (see Fig. \ref{extinct}). Other color correction terms were calculated, applying the formulas given in
Rieke \& Lebofsky (1985). {The ratio of total to selective extinction is $R_V=A_V/E_{B-V}=3.1$.} For reference for the SpT we used the table given in Kenyon \& Hartmann (1995) in the range from B0 to M6 and from M6.5 to L5 we used the reference stars collected {at web page of Neill Reid\footnote{http://www-int.stsci.edu/~inr/intrins.html}}. 

\begin{figure}
\resizebox{\hsize}{!}{\includegraphics{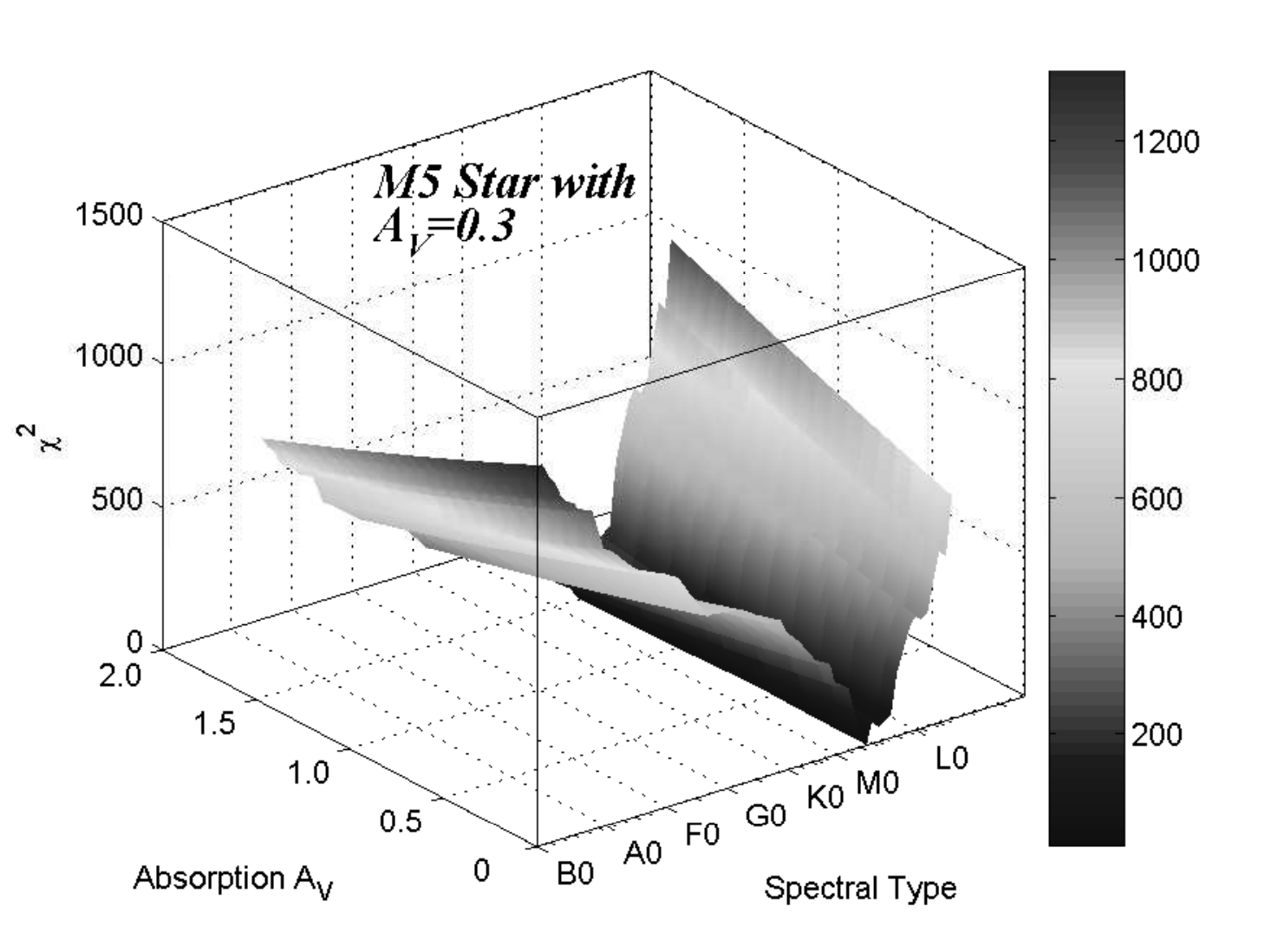}}
\caption{Example illustrating in which way we estimated the spectral type and extinction according to the $RIJHK_s$ photometry. For an assumed extinction the $\chi^2$ minimization is used to fit the data to the corresponding spectral type, defined by the same colors by tables of Kenyon \& Hartmann (1995) and Reid, N. The extinction is varied within a reasonable range ($A_V$ between 0 and 2 in this case) and a 2-dimensional grid of $\chi^2$ is created. The global minimum of this grid gives the extinction and the spectral types. In the tables \ref{tab1} and \ref{tab2} the uncertainties of the spectral type and the extinction according to the 1-$\sigma$ error of $\chi^2$ is given. }
\label{chisquare}
\end{figure}

Given these data we performed a $\chi^2$ minimization for each object with reference to the given information. Within the SpT vs. Av plane a global minimum in $\chi^2$ should indicate the most probable color at the most probable $A_V$. An example for the visualisation is given in Fig. \ref{chisquare}. Given an assumed extinction the SpT of the object is determined by fitting the data in Kenyon \& Hartmann (1995) or the data of Reid, assuming that all objects are main sequence stars (i.e. luminosity class \rm{V} dwarfs). This is repeated for a range of $A_V$ creating a 2- dimensional grid of $\chi^2$. The global minimum of this grid indicates the best SpT and $A_V$. The error bars of the photometry are {taken into account during the minimization of $\chi^2$. The results of that effort are summarized in Tbl. \ref{tab1}. With the same method we studied also objects which are presumably not members of the Pleiades, but happen to lie within our FoV and have 2MASS counterparts. Their photometry and spectral type is given in Tbl. \ref{tab2}. To estimate the uncertainties of our resulting quantities (SpT and $A_V$) we normalized the $\chi^2$ matrix. The 1-$\sigma$ error range for the normalized $\chi^2$ of a 2-dimensional fit is $\pm 1$. Every SpT/$A_V$ combination having a normalized $\chi^2$ within this interval can be a valid result and determines the uncertainty of SpT and $A_V$ as given in the brackets in the corresponding rows of Tbl. \ref{tab1} and \ref{tab2}. In case of poor photometry it happens, that the $\chi^2$ minimization results in more than one (typically 2) local minima. Typically one of these minima is located at low extinction and a late spectral type, while the other minimum can be found at a high value of extinction with an early spectral type. We tried to minimize that effect by presetting the possible range of $A_V=0 \ldots 2$ (objects with dusty envelopes or distant galaxies are probably not well determined). However in this cases a wide range of possible spectral types and $A_V$s occurs in the tables and the $\chi^2$ is typically large, indicating that the fitting results are not unique.} Finally the objects without near infrared (NIR) counterpart (with more Pleiades membership candidates among them) are listed in Tbl. \ref{tab3}. The table is divided by horizontal lines to indicate again brown dwarf candidates, stellar Pleiades membership candidates and all other stars in that order. 

\begin{figure*}
\begin{center}
\resizebox{\hsize}{!}{\includegraphics{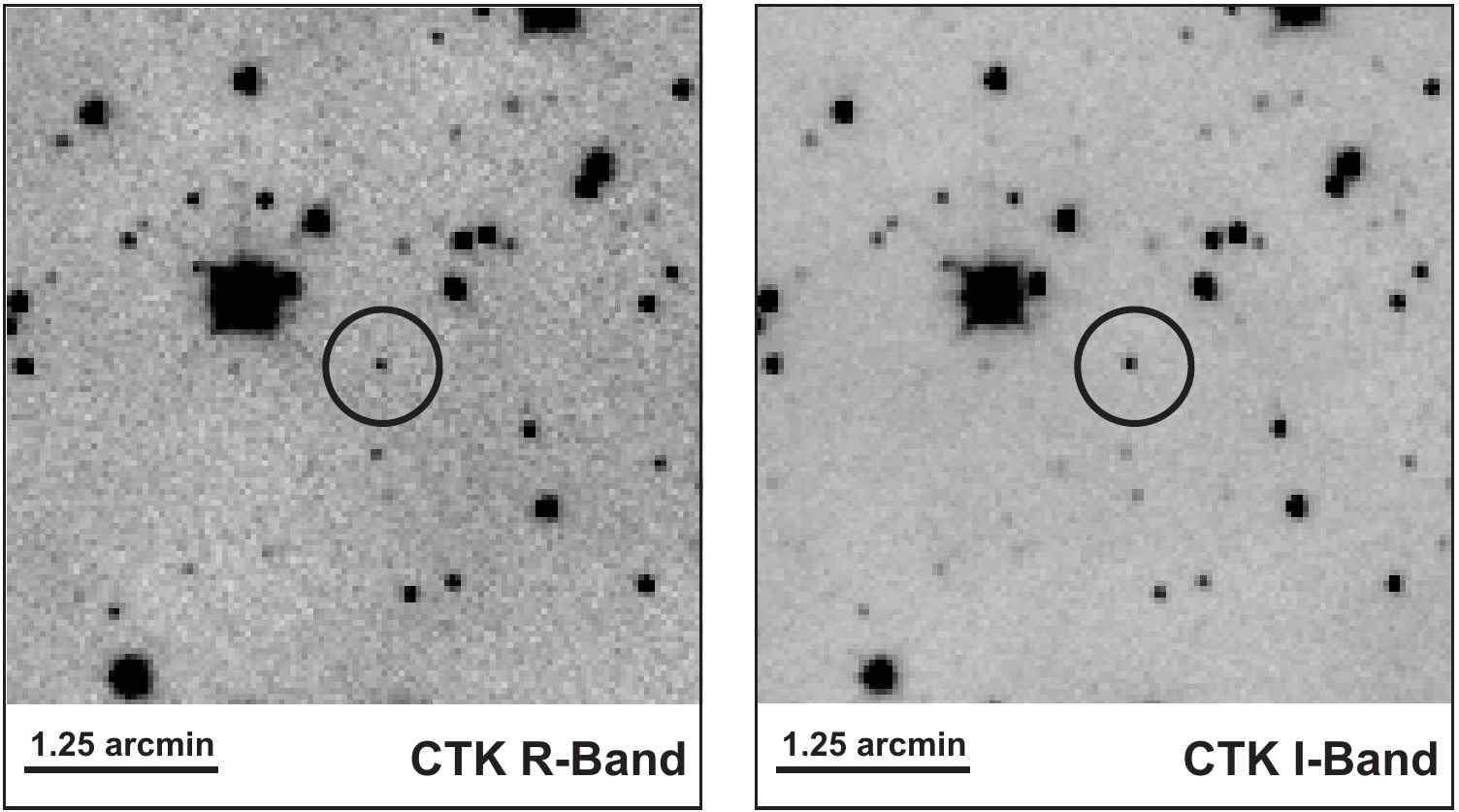}}
\end{center}
\caption{2MASS J3414296+2540432}
\label{cut3}
\end{figure*}

\begin{figure*}
\begin{center}
\resizebox{\hsize}{!}{\includegraphics{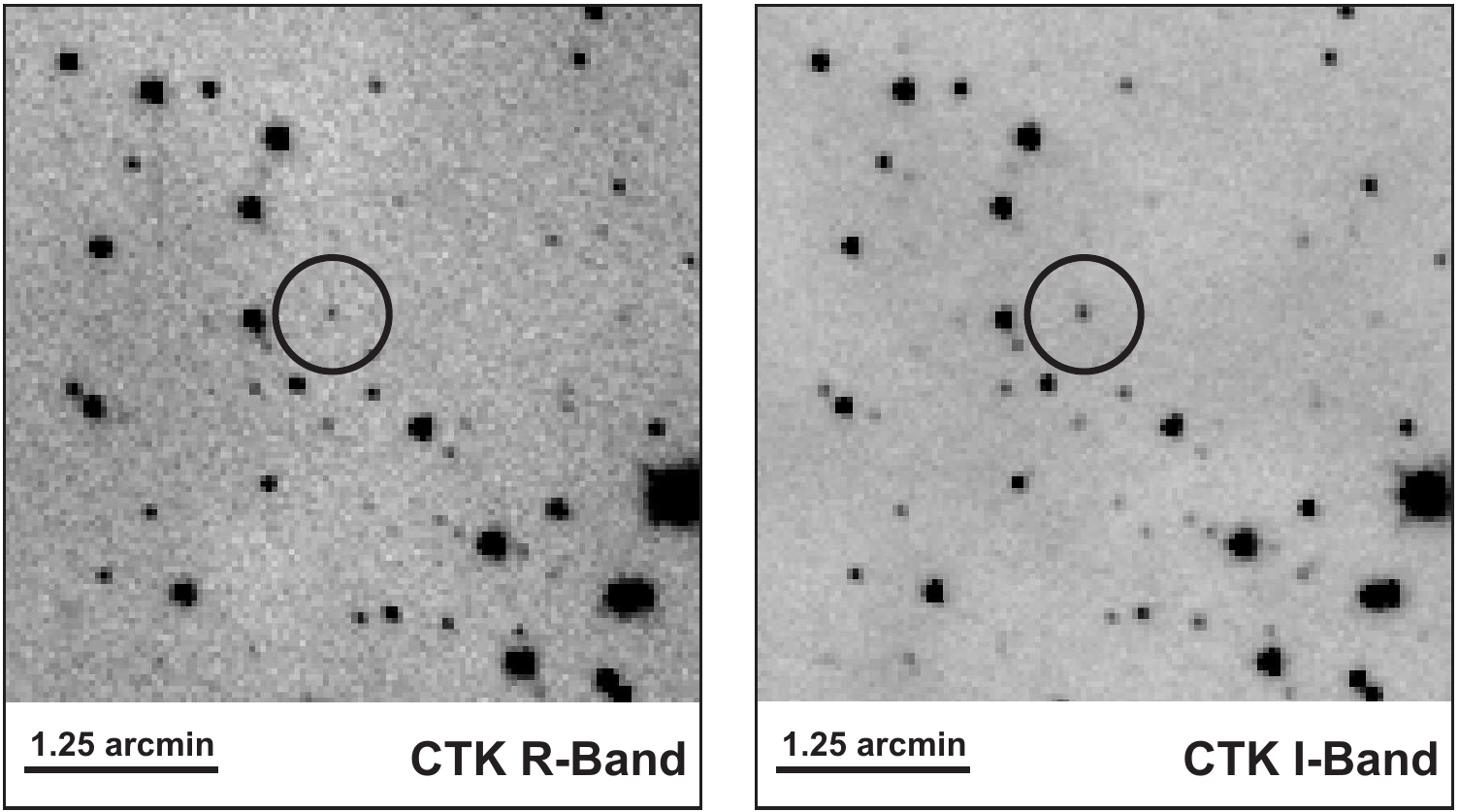}}
\end{center}
\caption{2MASS\,J3421030+2529316}
\label{cut4}
\end{figure*}

\begin{figure*}
\begin{center}
\resizebox{\hsize}{!}{\includegraphics{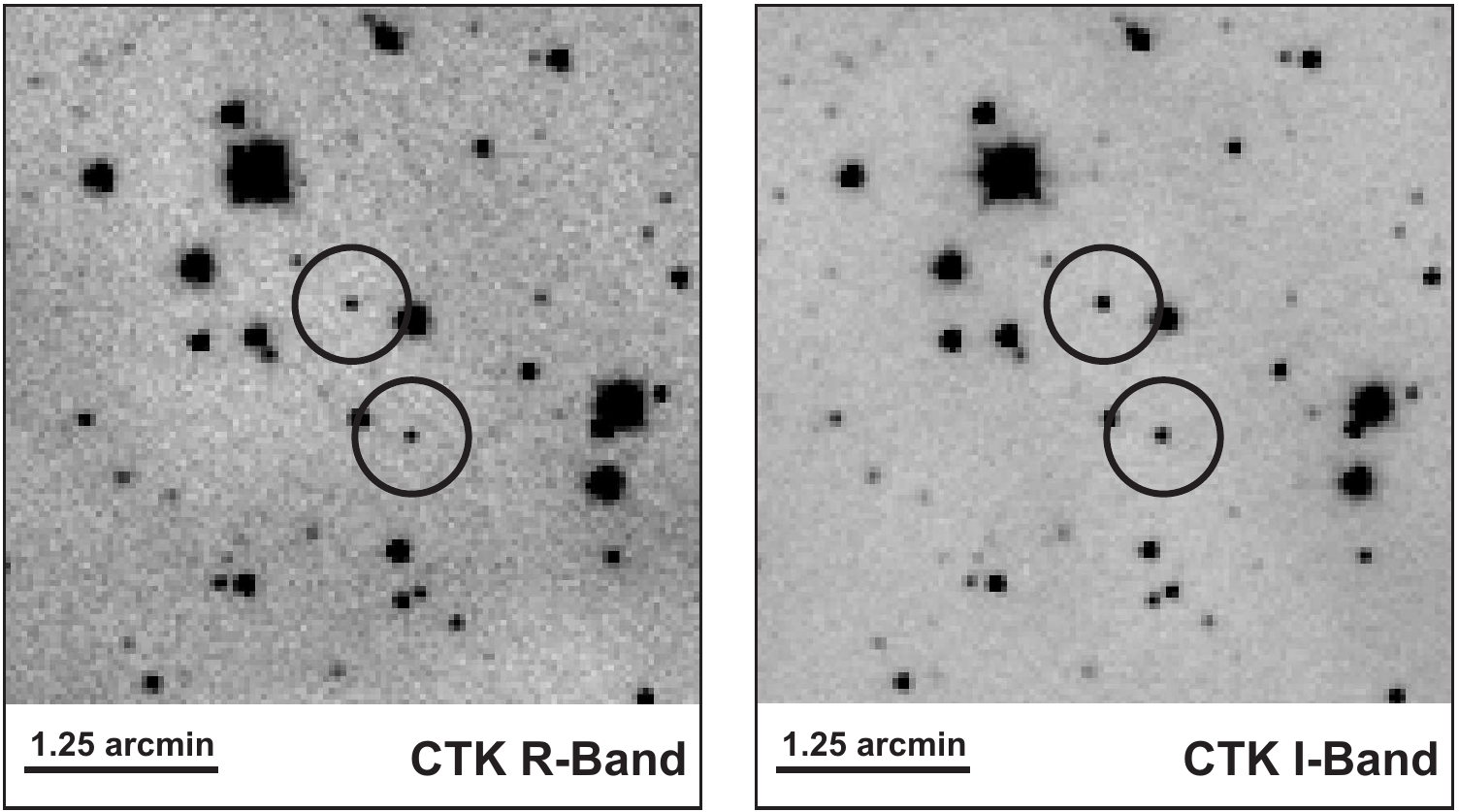}}
\end{center}
\caption{2MASS J3423655+2542193 (lower) and 2MASS J3423828+2543104 (upper)}
\label{cut5}
\end{figure*}

\begin{figure*}
\begin{center}
\resizebox{\hsize}{!}{\includegraphics{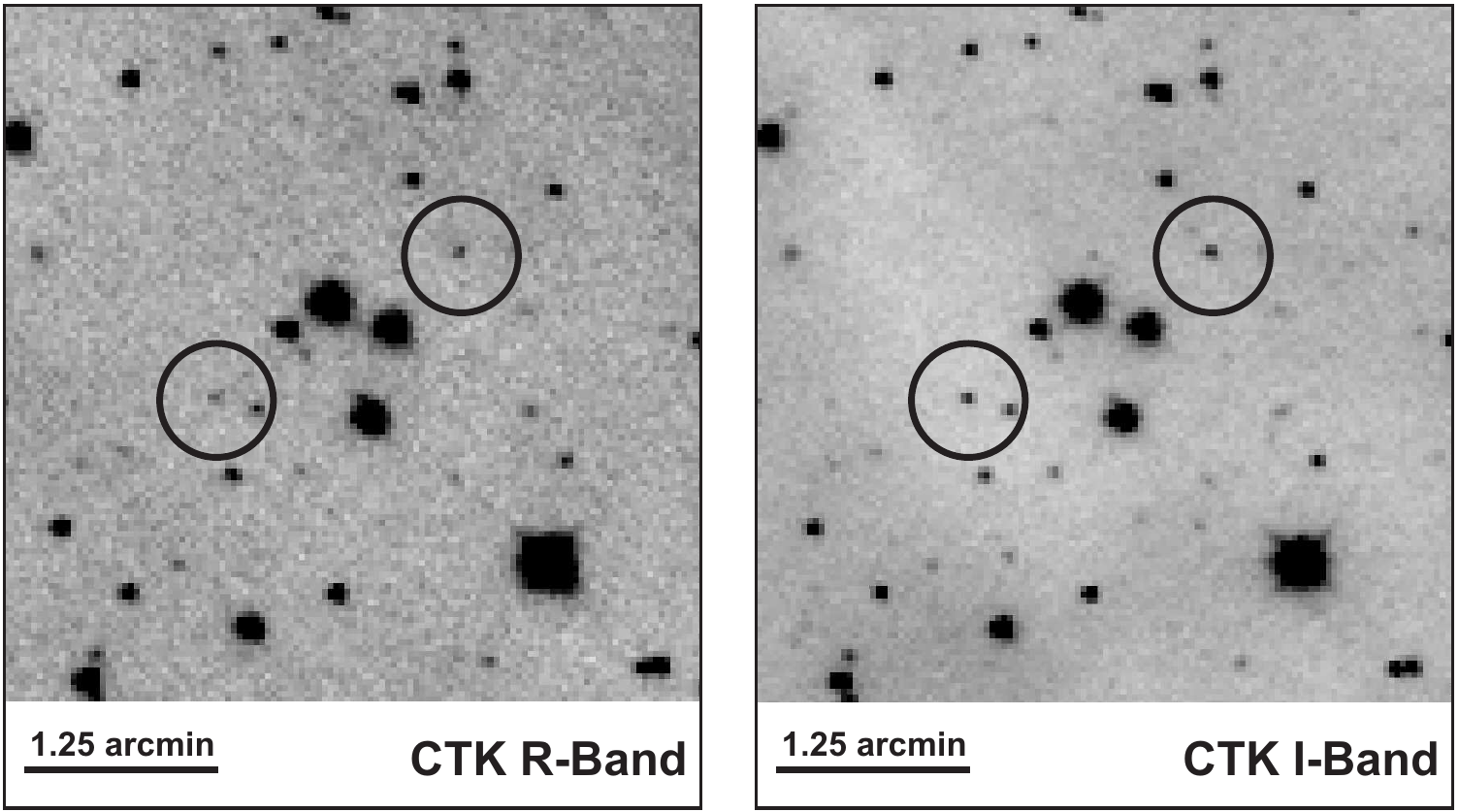}}
\end{center}
\caption{2MASS J3425334+2523044 (upper) and 2MASS J3430027+2522082 (lower)}
\label{cut7}
\end{figure*}

\begin{figure*}
\begin{center}
\resizebox{\hsize}{!}{\includegraphics{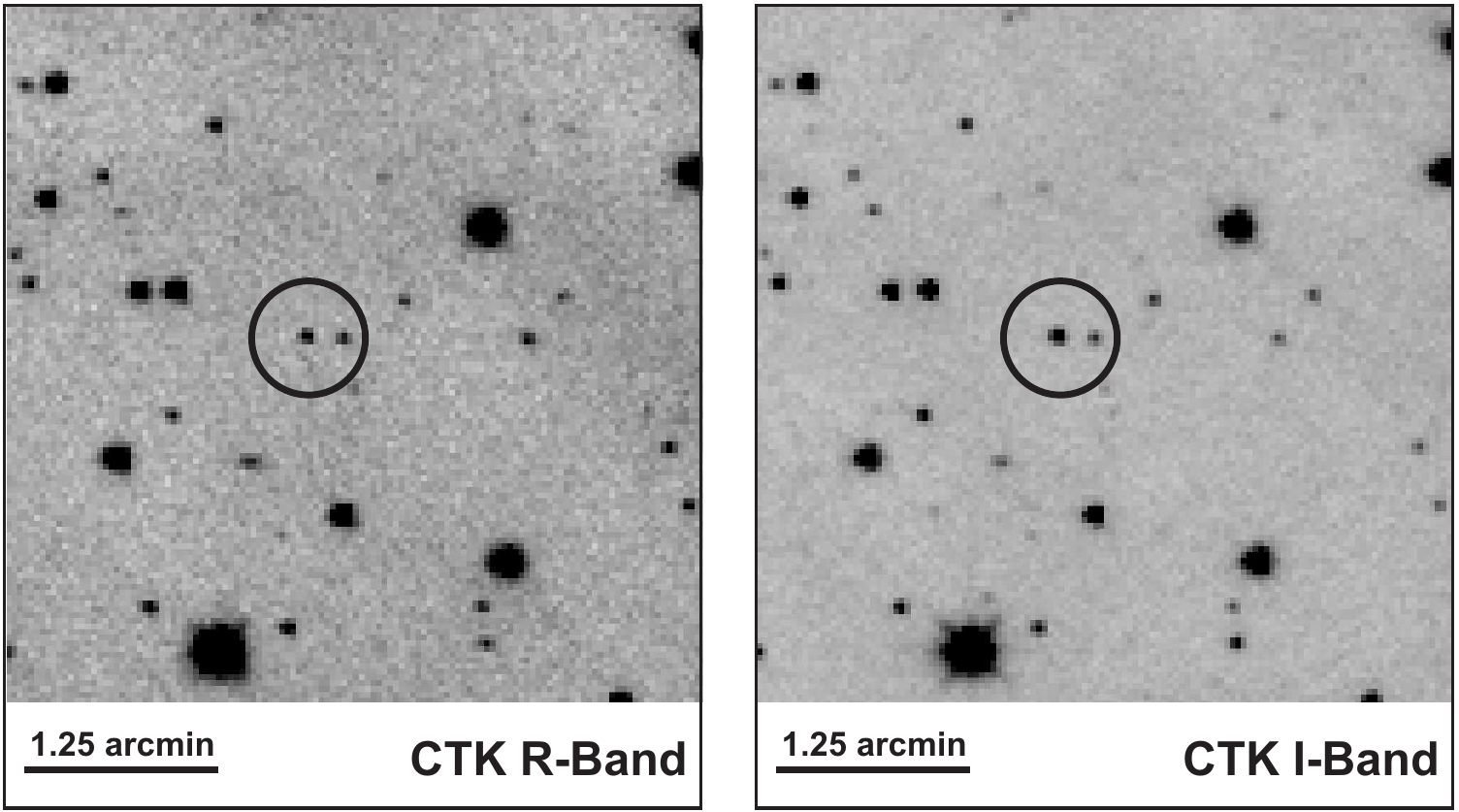}}
\end{center}
\caption{2MASS J3430237+2530225}
\label{cut8}
\end{figure*}

Out of the 197 candidate members 104 are found in the 2MASS catalog. Although these are the bright objects 7 of them are brown dwarf candidates. The individual objects are shown in Fig. \ref{cut3} to Fig. \ref{cut8}. The border between BDs and LMSs was defined according to the most massive BD known in the Pleiades, Teide 2 (Mart\'in \& Basri 1998), which has a spectral type of M6.5, a mass of $0.072\,M_{\odot}$ and $I_c=17.82\,$mag and is considered to be an object on the LMS/BD border. The membership candidates  are listed in Tbl. \ref{tab1} (BD candidates first).

\begin{figure}
\resizebox{\hsize}{!}{\includegraphics{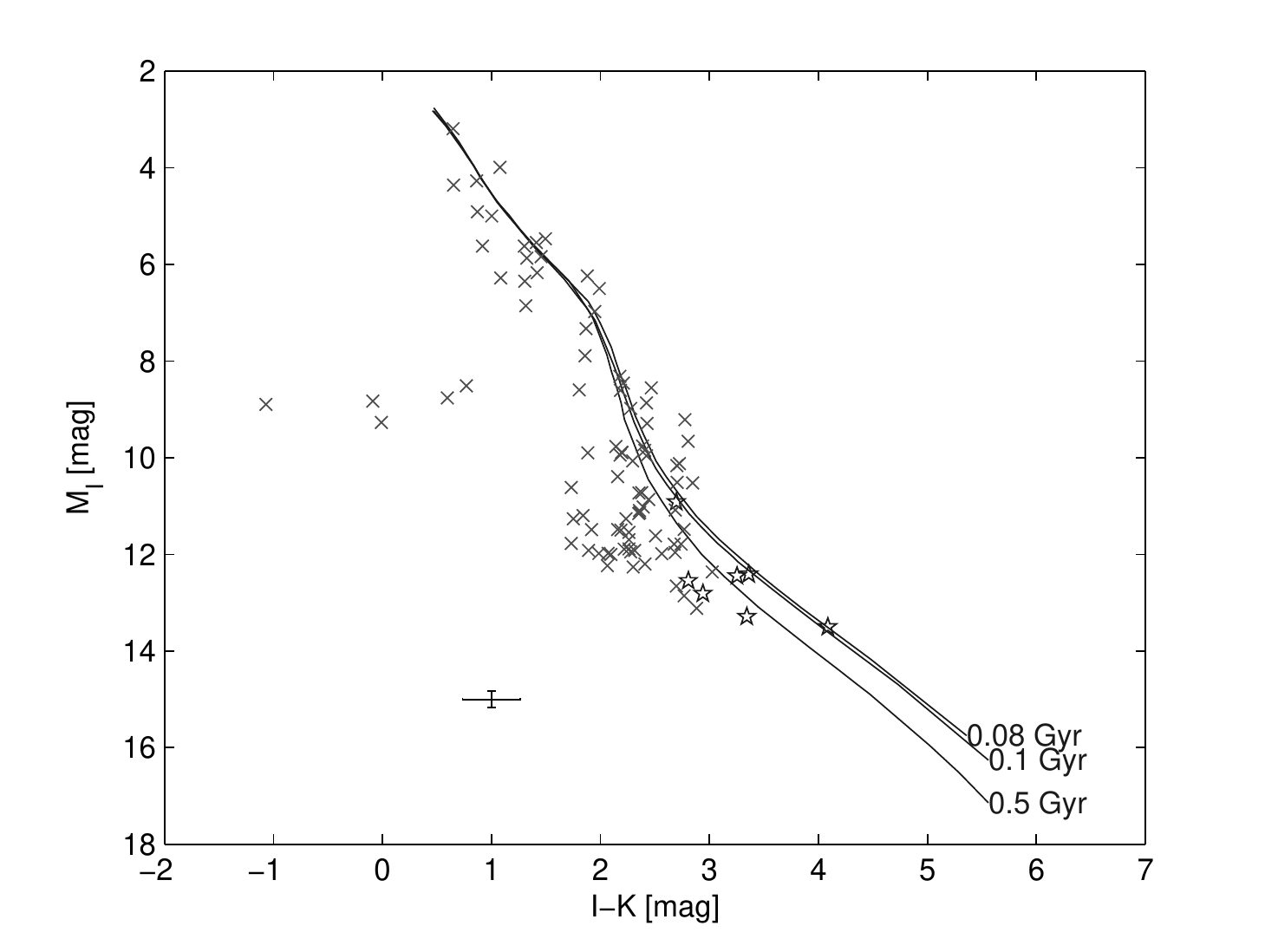}}
\caption{Comparison of the candidate members of the Pleiades (x) with Baraffe theoretical isochrones (Baraffe et al. 2002). A distance module of 5.66 as reported by An et al. (2007) is applied. Ages the of Isochrones are given in the figure and selected according to the reported age of the Pleiades of $\sim 120\,Myr$. In contrast to Fig. \ref{rimod} in the $I-K$ vs. $I$ plane the data of the brown dwarf candidates (stars) are in good  agreement with the Baraffe models while some other sources are not fitted at all by the isochrones. This indicates poor photometry for these objects or the assumption of Pleiades membership is wrong. For further explanations see text.}
\label{ikmod}
\end{figure}

{Fig. \ref{ikmod} shows the candidate members of the Pleiades in the $I-K$ vs. $I$ color magnitude diagram, together with the Baraffe isochrones already mentioned above and shown in Fig. \ref{rimod} in the $R-I$ vs. $I$ plane. The situation changed between the two images. Although in both images extinction corrected magnitudes are shown some objects do not fit the Baraffe isochrones. These objects do not have a fully consistent set of color information. The photometry of this sources is poor in some cases (see Tbl. \ref{tab1})}, for others the applied distance module is wrong since these are not Pleiades members. Furthermore the modeling might be inaccurate since it is still difficult to calculate the evolution of cool stars.  


\section{Summary and Conclusions}\label{concl}
We have obtained deep imaging observations of one specific Pleiades field, located at the rim of the Pleiades cluster with a size of $\sim 38'\times38'$. We have detected objects down to $R\sim 22\,$mag and $I\sim 20\,$mag. 
Among those objects 104 Pleiades candidates could be studied further since 2MASS photometry was available. This study identified 7 very red candidate members of the Pleiades. If these objects have Pleiades age and distance they could be massive brown dwarfs. This has to be ensured by spectroscopy and proper motion analysis, both not possible with the instrument used for this study. 

At the edge of a stellar cluster the member density is low. Such studies can give hints for BD formation theories. If BDs form like normal stars and high mass stars are formed in the center of molecular clouds it can be considered that more LMS and BD form further outside at the edges of clouds, where the material density is lower. Also BDs and LMS formed near the cluster center may get ejected and are now located near the edge or outside the cluster. {The frequency of BDs at the cluster edge should be different from the cluster center (see Kroupa \& Bouvier, 2003 and Kroupa et al. 2003 for a detailed discussion on this argument).} 


Further investigations are needed to confirm or reject the suspected BD candidates. Therefore a proper motion analysis and more precise photometry as well as spectroscopy is necessary. In particular NIR photometry of the other suspected membership candidates (not detected by 2MASS) is needed. 
%

\begin{acknowledgements}
TE would like to thank Katharina Schreyer for her help and advise.
RN acknowledges general support from the German National Science 
Foundation
(Deutsche Forschungsgemeinschaft, DFG) in grants NE 515/13-1, 13-2, and 
23-1,
AK acknowledges support from DFG in grant KR 2164/8-1,
SR and MV acknowledge support from the EU in the FP6 MC ToK project 
MTKD-CT-2006-042514,
TOBS acknowledges support from the Evangelisches Studienwerk e.V. 
Villigst,
TR acknowledges support from DFG in grant NE 515/23-1,
TE and MH acknowledge partial support from DFG in the SFB TR-7 Gravitation 
Wave
Astronomy. M. Moualla acknowledges support from the Syrian government.
This research has made use of the VizieR catalogue access tool and the Simbad database,
both operated at the Observatoire Strasbourg, as well as of the WEBDA database, operated at the
Institute for Astronomy of the University of Vienna and data products from the Two Micron All Sky Survey, which is a joint project of the University of Massachusetts and the Infrared Processing and Analysis Center/California Institute of Technology, funded by the National Aeronautics and Space Administration and the National Science Foundation, and of the NASA/ IPAC Infrared Science Archive, which is operated by the Jet Propulsion Laboratory, California Institute of Technology, under contract with the National Aeronautics and Space Administration.

\end{acknowledgements}

{\onecolumn
{\scriptsize
\appendix
\section{Lists of objects}
\nopagebreak
\begin{longtable}{l c@{$\pm$}c c@{$\pm$}c c@{$\pm$}c c@{$\pm$}c c@{$\pm$}c c@{$\pm$}c c l@{$\ \ \ $}l}
\nopagebreak
\caption{Extinction corrected colors and spectral type for the new Pleiades membership candidates found in this study. Most columns are self explaining. The {minimum} $\chi^2$ value for the estimated spectral type and interstellar extinction is given and can be seen as a quality factor. Objects above the first horizontal line have spectral types later or approximately M6.5 and so are brown dwarf candidates if they are Pleiades members. {Objects below the second horizontal line (at the end of the table) are sources located above the Pleiades main sequence in the color magnitude diagram. It is possible that these objects are unresolved binaries and they appear brighter in the sky (about 0.75 magnitudes in the worst case of equal mass binaries). Taking this into account these sources should be mentioned as additional candidate members of the Pleiades but the binary assumption has no observational evidence and the hypothesis of background giants can not be ruled out.}}\\
\hline\hline
         & \multicolumn{4}{|c|}{measured magnitudes}            & \multicolumn{8}{|c|}{extinction corrected colors} & & &\\
2MASS-ID & \multicolumn{2}{|c}{Rmag} & \multicolumn{2}{c|}{Imag} & \multicolumn{2}{|c}{R-I} & \multicolumn{2}{c}{I-J} & \multicolumn{2}{c}{I-H} & \multicolumn{2}{c|}{I-K$_s$} & $\chi^2$ & SpT & Av\\ 
\hline
\endfirsthead
\multicolumn{14}{l}{Table \ref{tab1} continued \ldots}\\
\hline
         & \multicolumn{4}{|c|}{measured magnitudes}            & \multicolumn{8}{|c|}{extinction corrected colors} & & &\\
2MASS-ID & \multicolumn{2}{|c}{Rmag} & \multicolumn{2}{c|}{Imag} & \multicolumn{2}{|c}{R-I} & \multicolumn{2}{c}{I-J} & \multicolumn{2}{c}{I-H} & \multicolumn{2}{c|}{I-K$_s$} & $\chi^2$ & SpT & Av\\ 
\hline
\endhead
J3414296+2540432 & 21.52& 0.88& 19.16 & 0.23& 2.36 & 0.91& 3.00 & 0.25& 3.58 & 0.25& 4.09 &0.26&  1.5 & M8 (M7.5..M8) & 0.00 (0.00..1.34) \\
J3421030+2529316 & 21.45& 0.77& 18.76 & 0.20& 2.53 & 0.79& 2.23 & 0.23& 2.29 & 0.29& 2.94 &0.27&  4.2 & M6 (M5..M7) & 0.61 (0.01..1.40) \\
J3423655+2542193 & 20.49& 0.39& 18.13 & 0.18& 2.36 & 0.43& 2.28 & 0.20& 3.17 & 0.21& 3.25 &0.21&  3.1 & M6.5 (M6..M7) & 0.04 (0.04..1.24) \\
J3423828+2543104 & 20.57& 0.41& 18.07 & 0.18& 2.50 & 0.44& 2.55 & 0.19& 3.13 & 0.20& 3.36 &0.20&  1.7 & M7.5 (M6.5..M7.5) & 0.00 (0.00..0.58) \\
J3425334+2523044 & 22.54& 1.82& 19.28 & 0.25& 3.07 & 1.83& 2.68 & 0.28& 2.03 & \ldots& 3.34 &0.32&  1.7 & M7.5 (M6.5..M7.5) & 0.69 (0.05..1.45) \\
J3430027+2522082 & 21.66& 0.87& 18.79 & 0.22& 2.55 & 0.90& 2.10 & 0.24& 2.59 & 0.27& 2.81 &0.30&  2.1 & M6 (M6..M7) & 1.20 (0.08..1.41) \\
J3430237+2530225 & 19.47& 0.28& 17.31 & 0.17& 1.76 & 0.33& 1.97 & 0.18& 2.48 & 0.18& 2.70 &0.18&  3.0 & M5 (M5..M6.5) & 1.53 (0.03..1.53) \\
\hline
J3410382+2532036 & 18.96& 0.32& 17.14 & 0.17& 1.63 & 0.36& 1.53 & 0.18& 2.16 & 0.18& 2.36 &0.18&  1.1 & M4 (M3..M4) & 0.72 (0.48..1.40) \\
\end{longtable}

{\scriptsize
\begin{longtable}{l c@{$\pm$}c c@{$\pm$}c c@{$\pm$}c c@{$\pm$}c c@{$\pm$}c c@{$\pm$}c c l@{$\ \ \ $} l}
\caption{Extinction corrected photometry and spectral type for objects not consistent with the Pleiades sample, but listed in the 2MASS catalog. These objects are most likely background objects - stars or distant galaxies. {As described in the text the grid of possible extinctions for the $\chi^2$ minimization is predefined $A_V=[0 \ldots 2]$. According to the limiting magnitude of the 2mass catalog ($K_s\approx15$) and of our survey most objects with 2mass counterpart (listed in this table) are red objects. Bluer objects without 2mass counterpart detected in this study are shown in Tbl. \ref{tab3}.} Some of the object are located near a saturated star (or are a saturated star itself), near the edge of the image, or near illumination effects visible in Fig. \ref{Rim} and Fig. \ref{Iim}. Typically this is indicated by a large value of $\chi^2$ and leads often to a wrong spectral type and unreliable extinctions. For the Pleiades membership candidates (Tbl. \ref{tab1} these objects are removed from the list but this is not taken into account for the rest of the sources. {Furthermore a wide range of possible spectral types indicates that the minimum found in the $\chi^2$ matrix is not unique, indicating that two different sets of SpT/Av values (late type star at low extinction, or early type star at high extinction) have nearly the same probability. As mentioned already in the text the spectral classification of such objects (often also indicated by a large $\chi^2$ value) is doubtful.}}\\
\hline\hline
         & \multicolumn{4}{|c|}{measured magnitudes}            & \multicolumn{8}{|c|}{extinction corrected colors} & & &\\
2MASS-ID & \multicolumn{2}{|c}{Rmag} & \multicolumn{2}{c|}{Imag} & \multicolumn{2}{|c}{R-I} & \multicolumn{2}{c}{I-J} & \multicolumn{2}{c}{I-H} & \multicolumn{2}{c|}{I-K$_s$} & $\chi^2$ & SpT & Av\\ 
\hline
\endfirsthead
\multicolumn{14}{l}{Table \ref{tab2} continued \ldots}\\
\hline
         & \multicolumn{4}{|c|}{measured magnitudes}            & \multicolumn{8}{|c|}{extinction corrected colors} & & &\\
2MASS-ID & \multicolumn{2}{|c}{Rmag} & \multicolumn{2}{c|}{Imag} & \multicolumn{2}{|c}{R-I} & \multicolumn{2}{c}{I-J} & \multicolumn{2}{c}{I-H} & \multicolumn{2}{c|}{I-K$_s$} & $\chi^2$ & SpT & Av\\ 
\hline
\endhead
J3405901+2545498 & 18.08& 0.27& 18.53 & 0.19& -0.82 & 0.33& 1.48 & 0.24& 1.92 & 0.27& 2.61 &0.24& 61.4 & K4 (G6..M2) & 1.39 (0.14..1.42) \\
J3405925+2526146 & 18.88& 0.28& 19.51 & 0.30& -1.14 & 0.41& 2.79 & 0.32& 3.14 & 0.34& 3.20 &0.35& 98.3 & M0 (G3..M5) & 1.90 (0.18..1.92) \\
J3405932+2523454 & 17.85& 0.27& 17.59 & 0.18& 0.15 & 0.32& 0.76 & 0.23& 1.61 & 0.23& 1.85 &0.24&  7.4 & K4 (G7..K6) & 0.40 (0.18..1.43) \\
J3405951+2524049 & 19.19& 0.32& 18.79 & 0.24& 0.38 & 0.40& 5.65 & 0.24& 6.16 & 0.24& 6.24 &0.24& 135.8 & L5 (M8..L5) & 0.04 (0.04..1.41) \\
J3405971+2534375 & 18.91& 0.29& 20.69 & 0.85& -1.89 & 0.90& 8.76 & 0.85& 9.07 & 0.85& 9.12 &0.85& 106.4 & L5 (M7.5..L5) & 0.42 (0.07..1.25) \\
J3410003+2546035 & 17.81& 0.27& 17.67 & 0.17& 0.15 & 0.32& 0.80 & 0.24& 1.08 & 0.33& 1.58 &\ldots&  2.7 & K0 (F6..K2) & 0.00 (0.00..0.72) \\
J3410029+2548560 & 16.27& 0.26& 14.80 & 0.17& 1.33 & 0.31& -1.87 & 0.21& -1.61 & 0.28& -1.07 &0.23& 102.5 & B0 (B0..F0) & 0.52 (0.21..0.93) \\
J3410058+2535137 & 18.35& 0.27& 18.04 & 0.18& 0.03 & 0.33& 1.27 & 0.21& 1.67 & \ldots& 1.65 &0.30&  8.9 & K4 (G9..K7) & 1.05 (0.02..1.48) \\
\end{longtable}

{\scriptsize
\begin{longtable}{l c@{$\,\pm$\,}c c@{$\,\pm$\,}c |l c@{$\,\pm$\,}c c@{$\,\pm$\,}c}
\caption{Basic data of detected objects which where not detected by the 2MASS. The table is divided by two horizontal lines in three sections. First red objects with Pleiades like $RI$ photometry are listed. These are more Pleiades BD candidates. Second bluer objects with Pleiades like Photometry are listed and afterward again objects which fall onto the Pleiades main sequence assuming that they are unresolved equal mass binaries. All classes of objects are separated by horizontal lines. Last objects with apparent $R$ and $I$ magnitudes making it unlikely for them to be members of the Pleiades. All objects listed in this table are detected on both, $R$ and $I$ band images at the same position. So (almost) all objects contained in this list are real astronomical objects from outside our solar system. Since none of these objects is listed in the 2mass catalog most of them are blue.}\\
\hline\hline
Id & \multicolumn{2}{c}{Imag}& \multicolumn{2}{c|}{$R-I$}&  Id & \multicolumn{2}{c}{Imag}& \multicolumn{2}{c}{$R-I$}\\
\hline
\endfirsthead
\multicolumn{10}{l}{Table \ref{tab3} continued \ldots}\\
\hline
Id & \multicolumn{2}{c}{Imag}& \multicolumn{2}{c|}{$R-I$}&  Id & \multicolumn{2}{c}{Imag}& \multicolumn{2}{c}{$R-I$}\\
\hline
\endhead
GSHCTK J034105+253220 & 18.3 & 0.2 & 2.4 & 0.7 & GSHCTK J034107+254139 & 20.3 & 0.5 & 3.2 & 5.4\\
GSHCTK J034107+254707 & 20.3 & 0.5 & 2.4 & 2.8 & GSHCTK J034107+254006 & 20.1 & 0.4 & 4.0 & 8.9\\
GSHCTK J034108+252851 & 18.1 & 0.2 & 2.4 & 0.5 & GSHCTK J034109+252603 & 18.4 & 0.2 & 2.2 & 0.5\\
GSHCTK J034110+251957 & 19.3 & 0.2 & 2.4 & 1.0 & GSHCTK J034110+252158 & 18.9 & 0.2 & 2.3 & 0.9\\
GSHCTK J034111+253718 & 18.2 & 0.2 & 2.2 & 0.4 & GSHCTK J034111+253557 & 19.7 & 0.3 & 2.4 & 1.4\\
GSHCTK J034115+254657 & 19.6 & 0.3 & 2.3 & 1.2 & GSHCTK J034116+252519 & 17.8 & 0.2 & 2.4 & 0.4\\
GSHCTK J034116+253444 & 19.1 & 0.2 & 2.4 & 0.8 & GSHCTK J034116+254743 & 18.7 & 0.2 & 2.7 & 0.8\\
GSHCTK J034117+252526 & 17.9 & 0.2 & 2.1 & 0.4 & GSHCTK J034117+253309 & 19.5 & 0.3 & 2.3 & 1.0\\
\end{longtable}}
}


\begin{thebibliography}{}
\bibitem{} An, D., Terndrup, D.M., Pinsonneault, M.H., et al\ 2007, ApJ~655, 233
\bibitem{} Banse, K., Ounnas, C., Ponz, D., Grosbol, P., \& Warmels, R.\ 1987, BA\&AS~19, 738 
{\bibitem{} Baraffe, I., Chabrier, G., Allard, F., \& Hauschildt, P.~H.\ 2002, A\&A~382, 563} 
\bibitem{} Bertin, E., \& Arnouts, S.\ 1996, A\&APS~117, 393
\bibitem{} Basri, G., Marcy, G., \& Graham, J.R.\ 1996, ApJ~458, 600 (BMG)
\bibitem{} Bertin, E., Mellier, Y., Radovich, M., et al.\ 2002 ASPC~281, 228
\bibitem{} Bertin, E.\ 2006, ASPC~351, 112
\bibitem{} Bouvier, J.R., Stauffer, E.L., Mart\'in, D., et al.\ 1998, A\&A~336, 490
\bibitem{} Cutri, R.M., Skrutskie, M.F., van Dyk, S., et al.\ 2003, 2MASS All
  Sky Catalog of point sources. (The IRSA 2MASS All-Sky Point Source Catalog,
  NASA/IPAC Infrared Science
  Archive.~http://irsa.ipac.caltech.edu/applications/Gator/)
 \bibitem{}Deacon, N.R., \& Hamply, N.C.\ 2004, VizieR On-line Data Catalog J/A+/A/416/125
 \bibitem{}Deacon, N.R., \& Hamply, N.C.\ 2004, A\&A~416, 125
 \bibitem{} Devillard, N.\ 2001, Astronomical Data Analysis Software and Systems X, 238, 525 
\bibitem{} Eisenbeiss, T., Seifahrt, A., Mugrauer, M., et al.\ 2007, AN~328, 521
\bibitem{} Kenyon, S.~J., \& Hartmann, L.\ 1995, ApJ~101, 117 
{\bibitem{} Kroupa, P., \& Bouvier, J.\ 2003, MNRAS~346, 369 
\bibitem{} Kroupa, P., Bouvier, J., Duch{\^e}ne, G., \& Moraux, E.\ 2003, MNRAS~346, 354}
\bibitem{} Mart\'in, D., Dahm, S.: 2001, ASPC~245, 349
\bibitem{} Mart\'in, E.L., \& Basri, G.\ 1998, ApJ~499, 64
\bibitem{} Mermilliod, J.-C.\ ed. 1998, WEB Acces to the Open Cluster Database
\bibitem{} Moraux, E., Bouvier, J.R., Stauffer, E.L., et al.\ 2003, A\&A~400, 891
\bibitem{} Moualla, M., Raetz, S., Neuh\"auser, R., et al.: in prep. 
\bibitem{} Mugrauer, M.\ 2009, AN, this issue
\bibitem{} Ochsenbein, F., Bauer, P., \& Marcout, J.\ 2000, A\&As~143, 23
\bibitem{} Rebollo, R., Mart\'in, E.L., Basri, G., et al.\ 1996, ApJS~469, L53
\bibitem{} Rieke, G.~H., \& Lebofsky, M.~J.\ 1985, ApJ~288, 618 
\bibitem{} Schwartz, M.J., \& Becklin, E.E.\ 2005, AJ~130, 2352
\bibitem{} Skrutskie, M.~F., et al.\ 2006, AJ~131, 1163 

\end{thebibliography}
\end{document}